\documentclass{LMCS}

\def\dOi{10(1:18)2014}
\lmcsheading%
{\dOi}
{1--21}
{}
{}
{Sep.~16, 2013}
{Mar.~26, 2014}
{}

\ACMCCS{[{\bf Mathematics of computing}]: Discrete
    mathematics---Graph theory---Graph Algorithms; [{\bf Theory of
        computation}]: Logic---Finite Model Theory}

\usepackage[utf8]{inputenc} 
\usepackage[T1]{fontenc}

\usepackage{enumerate} 
\usepackage{hyperref}

\usepackage{graphicx} 

\usepackage{url}

\begin{document}

\title[Model Checking Lower Bounds for Simple Graphs]{Model Checking
  Lower Bounds for Simple Graphs\rsuper*}

\author{Michael Lampis}
\address{Research Institute for Mathematical Sciences (RIMS), Kyoto University}
\email{mlampis@kurims.kyoto-u.ac.jp}
\thanks{This research was partially supported by the Scientific Grant-in-Aid
from  Ministry of Education, Culture, Sports, Science and Technology of Japan.
Part of this work was done while the author was at KTH Royal Institute of
Technology, supported by ERC Grant 226203}


\titlecomment{{\lsuper*}A conference version of this paper appeared in ICALP 2013.}

\begin{abstract}

A well-known result by Frick and Grohe shows that deciding FO logic on trees
involves a parameter dependence that is a tower of exponentials. Though this
lower bound is tight for Courcelle's theorem, it has been evaded by a series of
recent meta-theorems for other graph classes. Here we provide some additional
non-elementary lower bound results, which are in some sense stronger. Our goal
is to explain common traits in these recent meta-theorems and identify barriers
to further progress.

More specifically, first, we show that on the class of threshold graphs, and
therefore also on any union and complement-closed class, there is no
model-checking algorithm with elementary parameter dependence even for FO
logic. Second, we show that there is no model-checking algorithm with
elementary parameter dependence for MSO logic even restricted to paths (or
equivalently to unary strings), unless E=NE.  As a corollary, we resolve an
open problem on the complexity of MSO model-checking on graphs of bounded
max-leaf number. Finally, we look at MSO on the class of colored trees of depth
$d$. We show that, assuming the ETH, for every fixed $d\ge 1$ at least $d+1$
levels of exponentiation are necessary for this problem, thus showing that the
$(d+1)$-fold exponential algorithm recently given by Gajarsk{\'y} and
Hlin{\v{e}}n{\'y} is essentially optimal.

\end{abstract}

\maketitle

\section{Introduction}

Algorithmic meta-theorems are general statements establishing tractability for
a whole class of problems (often defined by expressibility in a certain logic)
in some class of inputs (usually a family of graphs). By far the most famous
and celebrated theorem in this area is a twenty-year old result due to
Courcelle \cite{Courcelle90} which states that all problems expressible in
monadic second-order logic (MSO$_2$) are linear-time solvable on graphs of
bounded treewidth. 
Thus, in one broad sweep this theorem establishes that a large number of
natural well-known problems, such as \textsc{3-Coloring} and
\textsc{Hamiltonicity}, are tractable on this important graph family. 
Much work has been devoted in recent years to proving stronger and stronger
meta-theorems in this spirit, often extending Courcelle's theorem to other
graph classes (see e.g.  \cite{CourcelleMR00,FrickG01,DawarGKS06} or
\cite{Grohe07,HlinenyOSG08} for some great surveys). 

The most often cited drawback of Courcelle's theorem has to do with the
``hidden constant'' in the algorithm's linear running time. It is clear that the
running time must somehow depend on the input formula and the graph's
treewidth, but the dependence given in Courcelle's theorem is in the worst case
a tower of exponentials whose height grows with the size of the formula.
Unfortunately, this cannot be avoided: Frick and Grohe \cite{FrickG04} proved
that the parameter dependence has to be non-elementary even if one restricts
the problem severely by just looking at properties expressible in first-order
logic on trees (unless P$=$NP). 

This lower bound result, though quite devastating, has proven very fruitful and
influential: several papers have appeared recently with the explicit aim of
proving meta-theorems which evade it, and thus achieve a much better dependence
on the parameters. Specifically, in \cite{Lampis12} algorithmic meta-theorems
with an elementary parameter dependence are shown for vertex cover, max-leaf
number and the newly defined neighborhood diversity. A meta-theorem for twin
cover was given by Ganian \cite{Ganian11}.  In addition, meta-theorems were
shown for tree-depth by Gajarsk{\'y} and Hlin{\v{e}}n{\'y} \cite{GajarskyH12}
and for the newly defined shrub-depth (which generalizes neighborhood diversity
and twin cover) by Ganian et al. \cite{GanianHNOMR12}. 

Thus, together with improved meta-theorems, these papers give a new crop of
graph complexity measures, some more general than others.  It becomes a natural
question how much progress we can hope to achieve this way, that is, how far
this process of defining more and more general ``graph widths'' can go on
before hitting some other natural barrier that precludes an elementary
parameter dependence.  Is simply avoiding the class of all trees enough?

This is exactly the question we try to answer in this paper. Towards this end
we try to give hardness results for graph families which are as simple as
possible.  Perhaps most striking among them is a result showing that not only
is avoiding all trees not enough but in fact it is necessary to avoid the much
smaller class of uncolored paths if one hopes for an elementary parameter
dependence.  As an example application, this almost immediately rules out the
existence of meta-theorems with elementary parameter dependence for any
induced-subgraph-closed graph class with unbounded diameter and any
edge-subdivision-closed graph class. This explains why all recently shown
meta-theorems we mentioned work on classes which are closed under induced
subgraphs but have bounded diameter and are not closed under edge subdivisions.

Our results can be summarized as follows. First, a non-elementary lower bound
for model checking FO logic on threshold graphs is shown. In a sense, this is a
natural analogue of the lower bound for trees to the realm of clique-width,
since threshold graphs are known to have the smallest possible clique-width.
The proof is relatively simple and consists mostly of translating a similar
lower bound given in \cite{FrickG04} for FO model checking on binary words.
However, the main interest of this result is that as a corollary we show that
the complexity of FO model checking is non-elementary for any graph class
closed under disjoint union and complement. This explains why, though some of
the recent meta-theorems work on complement-closed graph classes (e.g.
neighborhood diversity, shrub-depth) and some work on union-closed graph
classes (e.g.  tree-depth), no such meta-theorem has been shown for a class
that has both properties.

Our second result is that model checking MSO logic on uncolored paths (or
equivalently on unary strings) has a non-elementary parameter dependence. This
is the most technically demanding of the results of this paper, and it is
proved under the assumption E$\neq$NE. The proof consists of simulating the
workings of a non-deterministic Turing machine via an MSO formula on a path.
Though the idea of simulating Turing machines has appeared before in similar
contexts \cite{Kreutzer12}, because the graphs we have here are very restricted
we face a number of significant new challenges.  The main tool we use to
overcome them, which may be of independent interest, is an MSO formula
construction that compares the sizes of ordered sets while using an extremely
small number of quantifiers.  In the end, this result strengthens both
non-elementary MSO lower bounds given in \cite{FrickG04} (for trees and for
binary strings), modulo a slightly stronger complexity assumption. It also
resolves the complexity of MSO model checking for max-leaf number, which was
left open in \cite{Lampis12}. As an added corollary, we give an alternative,
self-contained proof of a result from \cite{CourcelleMR00}, stating that
MSO$_2$ model checking is not in XP for cliques unless E$=$NE. Furthermore,
we extend these ideas to order-invariant MSO logic, a more powerful variant of
MSO logic which has recently attracted interest
(\cite{EickmeyerKK13,EngelmannKS12}). We show that in this case the added power
is enough to make the parameter dependence non-elementary even in the most
trivial possible class of graphs, namely the class of edgeless graphs.

Finally, we study one of the recent positive results in this area by
considering the problem of model-checking MSO logic on rooted colored trees of
height $d$.  This is an especially interesting problem, since the $(d+1)$-fold
exponential algorithm of \cite{GajarskyH12} is the main tool used in the
meta-theorems of both \cite{GajarskyH12} and \cite{GanianHNOMR12}.  We show
that, assuming the Exponential-Time Hypothesis (ETH), any algorithm needs at
least $d+1$ levels of exponentiation, and therefore the algorithm of
\cite{GajarskyH12} is essentially optimal.  The main idea of the proof is to
``prune'' the trees constructed in the proof from \cite{FrickG04} and then use
an appropriate number of labels to differentiate their leaves.

\section{Preliminaries} \label{sec:defs}

The basic problem we are concerned with is model-checking: We are given a
formula $\phi$ (in some logic) and a structure $S$ (usually a graph or a
string) and must decide if $S\models \phi$, that is, if $S$ satisfies the
property described by $\phi$.  

We assume the reader is familiar with the basics of FO and MSO logic (see
e.g.  \cite{FrickG04}).  Let us just briefly describe some conventions.  We use
lower-case letters to denote vertex (FO) variables, and upper-case letters to denote
set variables. When the input is a graph, we assume the existence of an
$E(x,y)$ predicate encoding edges; when the input is a string a $\prec$
predicate encodes a total ordering; when the input is a rooted tree a $C(x,y)$
predicate encodes that $x$ is a child of $y$. Sometimes the input also has a
set of colors (also called labels). For each color $c$ we are given a unary
predicate $P_c(x)$. We will say that the input structure is a binary string
(that is, a string over the alphabet $\{0,1\}$) if the predicates given are
$\prec$ and $P_1$, where the $P_1$ predicate tells us which positions contain
the letter $1$.  To stress the distinction between set variables and the
supplied color predicates we will write $x\in S$ when $S$ is a set variable but
$P_i(x)$ where $P_i$ is a unary predicate supplied with the input.  We use
$x\preceq y$ as short-hand for $x\prec y \lor x=y$. To increase readability, we
also freely use basic set operations (e.g.~union, intersection, equality) with
the understanding that these can be implemented with standard MSO primitives in
a straightforward way.

When the input is an uncolored graph that consists of a single path it is
possible to simulate the $\prec$ predicate by picking one endpoint of the path
arbitrarily (call it $s$) and saying that $x\prec y$ if all paths from $s$ to
$y$ contain $x$. Thus, model-checking MSO logic on uncolored paths is at least
as hard as it is on unary strings. In most of the paper when we talk about MSO
logic for graphs we mean MSO$_1$, that is, with quantification over vertex sets
only. An exception is Corollary \ref{cor:cliques} which talks about MSO$_2$
logic, which allows edge set quantifiers.

We will also deal with order-invariant and successor-invariant versions of FO 
and MSO logic on graphs. Let $\phi$ be a logic formula using the $E(x,y)$
predicate and a binary predicate $<$. We say that $\phi$ is order-invariant if
for all graphs $G$ and for any two total orderings $<_1,<_2$ of the vertices of
$G$ we have that $(G,<_1)\models \phi$ if and only if $(G,<_2)\models \phi$.
Similarly, we say that a formula is successor-invariant if for any two
successor relations $S_1, S_2$ over the set of vertices we have $(G, S_1)
\models \phi$ if and only if $(G, S_2) \models \phi$. See
\cite{BenediktS05,EickmeyerKK13,EngelmannKS12} for more information on
order-invariant and successor-invariant logics.

A graph is a threshold graph (\cite{Chvatal77}) if it can be constructed from
$K_1$ by repeatedly adding \emph{union} vertices (not connected to any previous
vertex) and \emph{join} vertices (connected to all previous vertices), one at a
time.  Thus, a threshold graph can be described by a string over the alphabet
$\{u,j\}$. A graph is a cograph if it is $K_1$, or it is a disjoint union of
cographs, or it is the complement of a cograph. It is not hard to see that
threshold graphs are cographs. From the definition it follows that any class of
graphs that contains $K_1$ and is closed under disjoint union and complement
contains all cographs; if it is closed under the union and join operations it
contains all threshold graphs.

All logarithms are base two. We define $\exp^{(k)}(n)$ as follows:
$\exp^{(0)}(n)=n$ and $\exp^{(k+1)}(n)=2^{exp^{(k)}(n)}$. Then $\log^{(k)}n$ is
the inverse of $\exp^{(k)}(n)$. Finally, $\log^*n$ is the minimum $i$ such that
$\log^{(i)}n \le 1$. We use E (respectively NE) to denote that class of
problems decidable by a deterministic (respectively non-deterministic) Turing
machine in time $2^{O(n)}$.


\section{Threshold Graphs}

As mentioned, Frick and Grohe \cite{FrickG04} showed that there is no FPT
model-checking algorithm for FO logic on trees with an elementary dependence on
the formula size, under standard complexity assumptions. In many senses this is
a great lower bound result, because it matches the tower of exponentials that
appears in the running time of Courcelle's theorem, while looking both at a
much simpler logic (FO rather than MSO$_2$) and at the class of graphs with the
smallest possible treewidth, namely trees. 

Courcelle, Makowsky and Rotics \cite{CourcelleMR00} have given an extension of
Courcelle's theorem to MSO$_1$ logic for clique-width. The parameter dependence
is again a tower of exponentials and, since trees have cliquewidth at most 3
(\cite{CourcelleO00}), we already know that this cannot be avoided even for
graphs of constant clique-width. Here we will slightly strengthen this result,
showing that the non-elementary dependence cannot be avoided even on cographs,
the class of graphs that has the smallest possible clique-width (that is,
clique-width 2) without being trivial. We will heavily rely on a lower bound,
due again to Frick and Grohe \cite{FrickG04}, on the complexity of model
checking on binary strings.

One interesting consequence of the lower bound we give for cographs is that it
precludes the existence of an FPT algorithm with elementary parameter
dependence for any graph class that satisfies two simple properties: closure
under disjoint union and closure under complement. The reason for this is that
if a class is closed under both of these operations and it contains the
single-vertex graph, then it must contain all cographs (we will also show that
the assumption that $K_1$ is in the class is not needed).  This observation
helps to explain why, though some of the recent elementary model-checking
algorithms which have appeared work on union-closed graph classes, and some
work on complement-closed graph classes, no such algorithms are known for
classes with both properties.

The proof we present here is relatively simple and it relies on the following
theorem.

\begin{thm}[\cite{FrickG04}] 

Unless FPT=AW$[*]$, for any constant $c$ and any elementary function $f$ there
is no model-checking algorithm for FO logic on binary words which, given a
formula $\phi$ and a word $w$, decides if $w\models \phi$ in time at most
$f(\phi)|w|^c$.

\end{thm}

Let us recall that the assumption that FPT$\neq$AW$[*]$ is one of the standard
assumptions of parameterized complexity theory (\cite{flum2006,downeyf99}) and
is known to be weaker than the ETH.

We will reduce this problem to FO model checking on threshold graphs. This is
quite natural, since the definition of threshold graphs gives a straightforward
correspondence between graphs and strings.

\begin{thm} \label{thm:threshold}

Unless FPT=AW$[*]$, for any constant $c$ and any elementary function $f$ there
is no model-checking algorithm for FO logic on connected threshold graphs
which, given a formula $\phi$ and such a graph $G$, decides if $G\models \phi$
in time at most $f(\phi)|G|^c$.

\end{thm}

\proof

Suppose that we are given a binary word $w$ and an FO formula $\phi$. We will
reduce the problem of deciding if $w\models \phi$ to the problem of deciding if
$G\models\phi'$ for a threshold graph $G$ and a FO formula $\phi'$ which we
will construct.

First, let us describe $G$, and since it's a threshold graph we can describe it
as a string over the alphabet $\{u,j\}$. The graph $G$ starts with $uuj$. Then,
for each letter of $w$, if it is a 0 we append $uj$ to the description of $G$,
otherwise we append $ujj$. So, for example the graph corresponding to $w=01101$
would have description $uujujujjujjujujj$. Notice that, since the last letter
in the description is  $j$, the graph is connected.

Now we need to interpret the formula $\phi$ into the new context. To do this,
let's first observe some basic properties of our graph. First, a vertex in this
graph is a union vertex if and only if its neighborhood is a clique. To see
this, note that union vertices are only connected to join vertices, which, by
construction, form a clique. All join vertices on the other hand are connected
to the first two union vertices which are not mutually connected. Second, all
union vertices, except the first two (dummy) vertices have at least one join
vertex as a non-neighbor, namely at least the first join vertex.

We thus define the following formulas\vspace{-3 pt}

\begin{eqnarray*} 
union(x) &:=& \forall y\forall z\Big( (E(x,y)\land E(x,z)\land y\neq z)\to E(y,z)\Big) \\
main(x) &:=& union(x) \land \exists y \big(\neg union(y) \land \neg E(x,y) \big)
\end{eqnarray*}\vspace{-6 pt}

\noindent This will allow us to simulate selecting a letter in the
word by selecting the union vertex which represents the corresponding
pair or triple of vertices in the graph.

Now we also need to encode the $\prec$ and $P_1$ predicates. We define\vspace{-3 pt}

\begin{eqnarray*} 
prec(x,y) &:=& \exists z \big(\neg union(z) \land E(x,z) \land \neg E(y,z)\big)\\
one(x) &:=& \exists y \exists z\Big( (y\neq z) \land \neg union(y) \land \neg
union(z) \land E(x,y) \land E(x,z)  \\ 
&&\land \forall w \left(main(w) \land prec(x,w)\to\left(\neg E(y,w)\land\neg E(z,w)\right)\right)\Big)
\end{eqnarray*}\vspace{-6 pt}

\noindent The intuition for the first is that, if $x,y$ are two union
vertices that represent two different blocks, $x$ precedes $y$ if and
only if there exists some join vertex connected to $x$ but not
$y$. For $P_1$ we have that $x$ is a union vertex representing a $ujj$
block if and only if there exist two join vertices connected to it and
not connected to any union vertex that comes later in the description.

Given the above it is straightforward to produce the formula $\phi'$ from
$\phi$: the formulas $prec$ and $one$ are used to translate the corresponding
atomic predicates $\prec$ and $P_1$, while we inductively replace $\exists x
(\psi(x))$ with $\exists x (main(x)\land \psi'(x))$ where $\psi'(x)$ is the
translation of $\psi(x)$. It is not hard to see that $|\phi'|=O(\phi)$ while
the order of $G$ is $O(|w|)$. \qed

\begin{cor} \label{cor:union}

Let $\mathcal{C}$ be a non-empty graph class that is closed under disjoint
union and complement, or under disjoint union and join. Unless FPT=AW$[*]$, for
any constant $c$ and any elementary function $f$ there is no model-checking
algorithm for FO logic on $\mathcal{C}$ which, given a formula $\phi$ and a
graph $G\in\mathcal{C}$, decides if $G\models \phi$ in time at most
$f(\phi)|G|^c$.

\end{cor}

\proof

It suffices to prove this if the class is closed under union and join, because
if it's closed under union and complement we get closure under join ``for
free'' by successively performing a complement operation, followed by a
disjoint union, followed by a second complement operation.  The proof is
immediate if $K_1\in\mathcal{C}$, since then the class contains threshold
graphs.  Otherwise, let $G_m$ be the graph of the smallest order in the class
and say it has $k$ vertices. We will construct a graph as in the proof of
Theorem \ref{thm:threshold}, except that for each vertex we would be adding in
that case we will add a copy of $G_m$.  More specifically, for each union
vertex of the threshold graph we add a copy of $G_m$ to the graph we are
constructing without connecting it to any other vertex, and for each join
vertex we add a copy of $G_m$ and connect all its vertices to all previously
added vertices.  It's easy to see that the graph we have constructed is still
in $\mathcal{C}$.

It is now not hard to see how to translate the proof of Theorem
\ref{thm:threshold} in this case. Replace every $\exists x (\phi(x))$  with
$\exists x_1 \exists x_2\ldots \exists x_k (G_m(x_1,\ldots,x_k)\land
\phi(x_1))$, where $G_m$ is a formula stating that the $x_i$'s have the
structure of $G_m$ (that is, they are all distinct and have the same edges as
$G_m$) and they all have the same neighbors in the rest of the graph. We now
know that the $x_i$'s form a copy of $G_m$. If they all come from the vertices
added to represent a single letter, we then take one representative and use it
in the rest of the formula (note that if we take one representative from each
copy the result is a threshold graph). Observe that, if some of the $x_i$'s
come from two different copies of $G_m$, they must be two copies that were
added consecutively to represent letters of the same type (since the $x_i$'s
have the same neighbors in the rest of the graph). The remaining vertices of
the group that represents these letters also form a copy of $G_m$. Therefore,
there exists an automorphism which allows us, without loss of generality, to
assume that all the $x_i$'s correspond to a single letter.

This trick is sufficient to translate the formulas for $main$ and $prec$. The
only place where we may run into a problem are the $union$ and $one$ formulas,
because we use the $\neq$ predicate there. Since we are picking $x_1$ as an
arbitrary representative of a copy of $G_m$, if $G_m$ has a non-trivial
automorphism it could be the case that the $k$ vertices that correspond to
$\exists y$ and the $k$ vertices that correspond to $\exists z$ are assigned to
the same copy of $G_m$, but $y_1\neq z_1$. To avoid this case we just need to
add an extra formula after the quantification of $y,z$ stating that all
$y_i,z_j$ are pairwise distinct. \qed


\section{Paths, Unary Strings} \label{sec:paths}

The main result of this section is a reduction proving that, under the
assumption that E$\neq$NE, there is no FPT model-checking algorithm for MSO
logic with an elementary parameter dependence on the formula even on graphs
that consist of a single path, or equivalently, on unary strings.  As a
consequence, this settles the complexity of MSO model-checking on graphs with
bounded max-leaf number, a problem left open in \cite{Lampis12}, since paths
have the smallest possible max-leaf number (recall that a graph has max-leaf
number $k$ if it does not contain any tree with more than $k$ leaves as a
subgraph).  Until now a similar result was known only for the much richer class
of binary strings (or equivalently colored paths), under the weaker assumption
that P$\neq$NP \cite{FrickG04}. It is somewhat surprising that we are able to
extend this result to uncolored paths, because in this case the size of the
input is exponentially blown-up compared to a reasonable encoding.  One would
expect this to make the problem easier, but in fact, it only makes it more
complicated to establish hardness.

Indeed, one of the main hurdles in proving a lower bound for MSO on unary
strings, or paths, is information-theoretic. Normally, one would start with an
NP-hard problem, and reduce to a model-checking instance with a very small
formula $\phi$.  But, because the path we construct can naturally be stored
with a number of bits that is logarithmic in its size (by storing its length in
binary), in order to encode $n$ bits of information from the original instance
into the new instance we need to construct a path of exponential size.  Thus, a
polynomial-time reduction seems unlikely and this is the reason we end up using
the assumption that E$\neq$NE, instead of P$\neq$NP.

Our approach is to start from the prototypical problem for the class NE: given
$n$ bits of input for a non-deterministic Turing machine that runs in time
$2^{O(n)}$, does the machine accept? We will use the input path to simulate the
machine's tape and then ask for a subset of the vertices of this path that
corresponds to cells in the tape containing 1. Thus, what we need at this point
is an MSO formula that checks if the chosen vertices encode a correct accepting
computation.

Of course, to describe a machine's computation in MSO logic a significant
amount of machinery will be needed. We note that, though the approach of
simulating a Turing machine with an MSO formula has been used before (e.g.
\cite{Kreutzer12}), the problem here is significantly more challenging for two
reasons: first, unlike previous cases the input here is uncolored, so it is
harder to encode arbitrary bits; and second, there are (obviously) no grid-like
minors in our graph, so it's harder to encode the evolution of a machine's
tape, and in particular to identify vertices that correspond to the same tape
cell in different points in time.

Our main building block to overcome these problems is an MSO construction which
compares the sizes of paths (or generally, ordered sets) of size $n$ with very
few (roughly $2^{O(\log^*n)}$) quantifiers. We first describe how to build this
formula, then use it to obtain other basic arithmetic operations (such as
exponentiation and division) and finally explain how they all fit together to
give the promised result. This construction may be of independent interest in
the context of the counting power of MSO logic: recall that for unordered sets
an MSO formula with $q$ quantifiers cannot distinguish two sets of different
size if they are both of size larger than $2^q$ (this is the basis of the
results of \cite{Lampis12}). In contrast, our construction implies that for
ordered sets MSO formulas with $q$ quantifiers can distinguish sets whose sizes
are a non-elementary function of $q$.

\subsection{Measuring Long Paths with Few Quantifiers} 

To keep the presentation simple we will concentrate on the model-checking
problem on unary strings, that is, linearly ordered sets; formulas for MSO on
paths can easily be constructed as explained in section \ref{sec:defs}. We
therefore assume that there is a predicate $\prec$ which gives a total ordering
of all elements.

Let us now develop our basic tool, which will be an MSO formula
$eq_L(A_1,A_2)$, where $A_1,A_2$ are free set variables. The desired behavior
of the formula is that if $|A_1|=|A_2|$ and $|A_1|\le L$ then the formula will
be true, while on the other hand whenever the formula is true it must be the
case that $|A_1|=|A_2|$.  In other words, the formula will always correctly
identify equal sets with size up to $L$, and it will never identify two unequal
sets as equal (it may however be false for two equal sets larger than $L$).
Our main objective is to achieve this with as few quantifiers as possible.

We will work inductively. It should be clear that for very small values of $L$
(say $L=4$) it is possible to compare sets of elements with size at most $L$
with a constant number of set and vertex quantifiers and we can simply make the
formula false if one set has more than $4$ elements.  So, suppose that we have
a way to construct the desired formula for some $L$.  We will show how to use
it to make the formula $eq_{L'}$, where $L'\ge L\cdot 2^L$.  If our recursive
definition of $eq_{L'}$ uses a constant number of copies of $eq_L$ then in the
end we will have $|eq_L|=2^{O(\log^*L)}$, because for each level of
exponentiation we blow up the size of a formula by a constant factor.  This
will be sufficiently small to rule out a non-elementary parameter dependence.

\begin{figure}

\centering \includegraphics[width=0.8\textwidth]{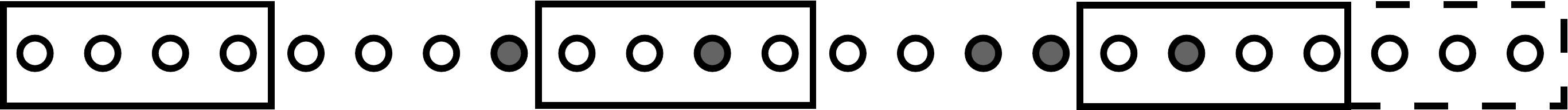}

\caption{An example of the counting structure imposed on a set for $L=4$.  We
select a set $Q$ that constructs equal-sized sections (denoted by boxes) and then a
second set $B$ that encodes a binary number in each section (shown in grey).}
\label{fig:counting}

\end{figure}

Let us now give a high-level description of the idea, by concentrating first on
the set $A_1$.  We will select a subset of $A_1$, call it $Q_1$, and this
naturally divides $A_1$ into sections, which are defined as maximal sets of
vertices of $A_1$, consecutive in the ordering, with the property that either
all or none of their vertices belong to $Q_1$.  We will make sure that all
sections have length $L$, except perhaps the last, which we call the remainder
(see Figure \ref{fig:counting}).  It is not hard to see that this structure can
be imposed if the predicate $eq_L$ is available. We do the same for $A_2$ and
now we need to verify that the two remainders have the same length (easy with
$eq_L$) and that we have the same number of sections in $A_1$ and $A_2$.

Verifying that we have the same number of sections is the interesting part. Now
we could naively try to count the number of sections by selecting a
representative from each and forming a set.  This would not work since the
number of sections is at most $2^L$ and the inductive hypothesis only allows us
to use $eq_L$ to compare sets of size $L$.  Thus, we have to work a little
harder.

We select another subset of $A_1$, call it $B_1$.  The intuition here is that
selecting $B_1$ corresponds to encoding a number in each section, by
interpreting elements that belong in $B_1$ as encoding 1 and the rest as 0. We
will now need to make sure that each section encodes the binary number that is
one larger than the number encoded by the immediately preceding section. This
is achievable by using $eq_L$ to locate the elements that represent the same
bit positions. We also make sure that there is no overflow in the counting and
that counting starts from zero, that is, all sections have some vertex not in
$B_1$ and the first has no vertices in $B_1$.

Finally, assuming that the above counting structure is correctly imposed on
both $A_1$ and $A_2$ all that is left is to take the last sections of both
$A_1$ and $A_2$ and compare them. If the same binary number is encoded in both
then $|A_1|=|A_2|$.

Let us now give a formal definition of $eq_L$. First, we need to be able to
recognize sections. Assume that we have a set of elements $U$ and a subset
$P\subseteq U$. Informally, $U$ is the set of elements we are currently working
on (so $U\subseteq A_1\cup A_2$). As explained $P$ divides $U$ into sections so
we define a formula $section(S,U,P)$ that will be true if $S$ is such a
section.\vspace{-3 pt} 

\begin{eqnarray*}
consec(S,U)&:=&\forall x\forall y\forall z \Big((x\in S) \land (y\in S) \land (z\in U)\land (x\prec z) \\
&& \ \land (z\prec y)\to z\in S \Big) \\
partsection(S,U,P) &:=& S\subseteq U \land consec(S,U)\land \\
&&\ \ \land \forall x\forall y \Big( (x \in S) \land (y\in S) \to  (x\in P \leftrightarrow y\in P)\Big)\\
section(S,U,P) &:=& partsection(S,U,P) \land\\
&&\ \ \forall S' (S\subseteq S' \land partsection(S',U,P) \to S'=S)
\end{eqnarray*}\vspace{-6 pt}

\noindent Informally, $consec$ checks if $S$ is a contiguous subset of
$U$. Then, $S$ is a partial section if it's a subset of either $P$ or
$U\setminus P$ and it represents a contiguous subset of $U$
elements. $S$ is a section if it's a maximal partial section.

Assuming two sets $S_1,S_2$ represent consecutive sections and we have a set
$B$ which is supposed to encode a binary number $i$ in $S_1$ and $i+1$ in $S_2$
we check this with the following formula (explained informally below):\vspace{-3 pt}

\begin{eqnarray*}
next(S_1,S_2,B) &:=& \exists S_1^L\exists s^1\exists S_1^R \exists S_2^L\exists s^2\exists S_2^R  \\
&& \bigwedge_{i=1,2} \left((S_i^L\subseteq S_i) \land (s^i\in S_i) \land (S_i^R\subseteq S_i)\right) \land \\
&& \bigwedge_{i=1,2} \left( \forall x\forall y ((x\in S_i^L)\land (y\in S_i^R) \to (x\prec s^i\land s^i\prec y)) \right) \land \\
&& \bigwedge_{i=1,2}	\left(S_i \subseteq (S_i^L\cup \{s^i\} \cup S_i^R) \right)\\
&&	\land\ (s^2\in B) \land (s^1\not\in B) \land (S_1^R\subseteq B) \land (S_2^R\cap B=\emptyset)\\
&& \land\ eq_L(S_1^R,S_2^R)\\
&&     \land\ same(S_1^L,S_2^L,B,B) \\
same(S_1,S_2,B_1,B_2)&:=& \exists f_1\exists f_2 \Big( \bigwedge_{i=1,2} \forall x (x\in S_i \to  f_i\preceq x ) \land \\
&& \forall S_1^L \forall S_2^L \Big(\bigwedge_{i=1,2} \big((f_i\in S_i^L) 
    \land consec(S_i^L,S_i)\big)\land eq_L(S_1^L,S_2^L) \to \\
&& \exists l_1\exists l_2 \Big(\bigwedge_{i=1,2} \forall x (x\in S_i \to  x \preceq l_i ) \land \\
&&\ \ \ \land (l_1\in B_1 \leftrightarrow l_2\in B_2)\Big)\Big) \Big)
\end{eqnarray*}\vspace{-6 pt}

\noindent Informally, $next$ partitions the sets $S_1,S_2$ into a left
and right part and identifies a vertex $s^i$ between the two
parts. Assuming $|S_1|=|S_2|$ respective parts have the same size; the
right part of $S_1$ corresponds to all 1 digits and the right part of
$S_2$ to all 0. The left parts have to encode the same number, which
is checked by $same$. The idea here is that if we select equal length
contiguous prefixes of the sets we are checking the last element will
encode the same digit in both. We allow $same$ to use different sets
$B_1,B_2$ to read the encoding in $S_1,S_2$. This extra generality
will be useful when we reuse this formula later.

We will use $next$ only for neighboring sections. To check if two disjoint
sections are indeed adjacent we define:\vspace{-3 pt}

\begin{eqnarray*}
adj(S_1,S_2,U) &:=& \exists x \exists y \Big(x\in S_1 \land y\in S_2 \land \forall z(z\in U \to (z\preceq x \lor y\preceq z))\Big)
\end{eqnarray*}\vspace{-3 pt}

\noindent Informally, $adj$ is true if $S_1$ is a section that
directly precedes the section $S_2$, because then $x$ is the last
element of $S_1$ and $y$ is the first element of $S_2$ and there are
no elements between them.

We are now ready to define $eq_{L'}$ for $L'=L\cdot 2^L$.\vspace{-3 pt}

\begin{eqnarray*}
eq_{L'}(A_1,A_2) &:=& \exists Q_1 \exists Q_2 \exists R_1 \exists R_2 \exists B_1 \exists B_2 \\
&&\Big(\bigwedge_{i=1,2} \big( Q_i\subseteq A_i \land R_i\subseteq A_i \land B_i\subseteq A_i \big) \land eq_L(R_1,R_2)\land\\
&& \forall S_1 \forall S_2 \Big(\big(\bigwedge_{i=1,2}section(S_i,A_i\setminus R_i,Q_i)\big)  \to eq_L(S_1,S_2)\Big)\land\\
&&\bigwedge_{i=1,2} \Big( \forall S\forall S' \big((S\cap S'=\emptyset) \land section(S,A_i\setminus R_i,Q_i) \land \\
&&\ \ section(S',A_i\setminus R_i,Q_i) \land\ adj(S,S',A_i\setminus R_i)\big) \\
&&\ \ \to next(S,S',B_i)\Big) \land \\
&&\exists S_1^F\exists S_2^F \Big( same(S_1^F,S_2^F,B_1,B_2)\land\\
&&\ \  \bigwedge_{i=1,2}\Big( section(S_i^F,A_i\setminus R_i,Q_i)\land\\
&&\ \ \ \ \ 	\forall S' \big(section(S',A_i\setminus R_i,Q_i) \land (S'\cap S_i^F=\emptyset) \\
&&\ \ \to\exists x\exists y(x\in S' \land y \in S_i^F \land x\prec y) \big)\Big)\Big) \land\\
&& \forall S \big(\bigwedge_{i=1,2} section(S,A_i\setminus R_i,Q_i) \to (S\setminus B_i\neq \emptyset)\big) \land \\
&& \bigwedge_{i=1,2} \exists S \big(section(S,A_i\setminus R_i,Q_i) \land (S\cap B_i= \emptyset)\big)  \Big)
\end{eqnarray*}\vspace{-6 pt}

\noindent This is rather long, so let us explain it intuitively. We
want to test if $|A_1|=|A_2|$, so we demand the following:

\begin{itemize}

\item From both we remove a remainder set $R_i$, and we make sure that the
remainder sets are of equal size.

\item We use $Q_1,Q_2$ to partition the two sets into sections. All sections of
the first set must be equal in size to all sections of the second (therefore,
all sections in both sets are equal).

\item Select the sets $B_i$ which will encode binary numbers in the sections.
For each two disjoint sections which are consecutive check that they encode
consecutive numbers.

\item Find the last section on each set ($S_1^F, S_2^F$). Check that they
encode the same number.

\item Check that no section encodes a number made up only of 1s, so we don't
have an overflow in the counting.

\item Check that the first section on each set encodes the number zero (that
is, it has no elements from $B_i$). For this it's sufficient to check that some
section encodes zero, since we have already established proper ordering.

\end{itemize}

\begin{lem} \label{lem:eqL}

Let $L>2$ be a power of two. Then we can define a formula $eq_L(A_1,A_2)$ such
that if $|A_1|=|A_2|<L\cdot \log L$ then the formula is true and also if the
formula is true then $|A_1|=|A_2|$. Furthermore $|eq_L|=2^{O(\log^*L)}$. 

\end{lem}

\proof

Correctness follows by induction and the definition of the construction given
above. For the size bound, note that for $L'=L\cdot 2^L$ we have
$|eq_{L'}|=O(|eq_L|)$, since the definition of $eq_{L'}$ uses $eq_L$ a constant
number of times. It follows that there exists a constant $c$ such that for all
$k$ we have $|eq_{exp^{(k)}(1)}|=O( c^k) $. The result follows by setting
$k=\log^*L$. \qed

\medskip

\begin{figure}

\centering \includegraphics[width=0.8\textwidth]{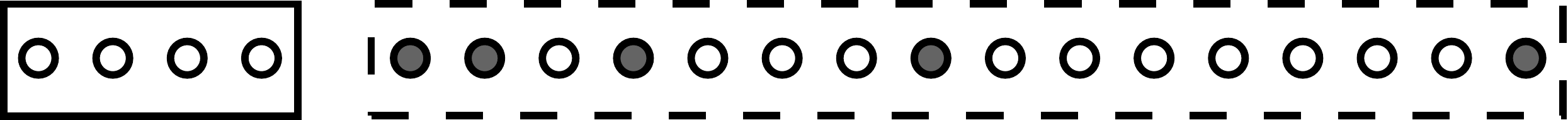}

\caption{An example where the set on the left has size $4$ and we verify that
the set on the right has size $2^4$. First we select a set of elements $Q$ on
the right of size one more than the size of the set on the left. Then we ensure
that distances between consecutive elements of $Q$ form a geometric progression.} \label{fig:pathexp}

\end{figure}

\noindent \textbf{Other arithmetic operations:} Before we go on, we will also
need formulas to perform some slightly more complicated arithmetic operations
than simply counting. In particular, we will need a formula $exp_L(A_1,A_2)$,
which will be true if $|A_2|=2^{|A_1|}$, assuming neither set has size more
than $L$.  The trick we use for this is shown in Figure \ref{fig:pathexp}.  The
idea is that we select a subset of $A_2$, call it $Q$, which marks out a set of
$|A_1|+1$ elements whose consecutive distances form a geometric progression
with ratio 2.

Let us now give some details. Again, we first define some auxiliary formulas.
Checking if a set is twice as large as another can be done as follows:\vspace{-3 pt}

\begin{eqnarray*} double(S_1,S_2) &:=& \exists S' \Big((S'\subseteq S_2) \land
eq_L(S_1,S') \land eq_L(S_1,S_2\setminus S')\Big) \end{eqnarray*}\vspace{-3 pt}

\noindent If we are given three elements $x,y,z$ such that $x\prec
y\prec z$ we can check that consecutive distances are doubled as
follows:\vspace{-3 pt}

\begin{eqnarray*} 
ddist(x,y,z,U) &:=& \exists S_1\Big( S_1\subseteq U\land  \forall u \left((x\preceq u \land u\prec y) \leftrightarrow u\in S_1 \right)\land\\
&&  \exists S_2 \Big(S_2 \subseteq U\land \forall u \left((y\preceq u \land u\prec z) \leftrightarrow u\in S_2\right)\\
&& \land\ \  double(S_1,S_2) \Big) \Big)
\end{eqnarray*}\vspace{-6 pt}

\noindent Informally, we select the set $S_1$ as the set of elements
starting from $x$ and up to (but not including) $y$, and $S_2$ as the
set of elements starting from $y$ and up to (but not including)
$z$. The second set must be twice as large.  Now we can define
$exp$:\vspace{-3 pt}

\begin{eqnarray*}
exp(A_1,A_2) &:=& \exists Q \exists f\exists s \exists l\Big( (Q\subseteq A_2) \land f\in Q \land s\in Q\land l\in Q \land f\prec s \land \\
&& \forall u \big(u\in A_2 \to (u\preceq l  \land (s\preceq u  \lor f=u))\big) \land \\
&& eq_L(A_1,Q\setminus\{l\}) \land \\
&&\forall x\forall y \forall z\big( x\in Q\land y\in Q\land z\in Q \land x\prec y \land y\prec z \land \\
&& \neg \exists u \big( u\in Q \land (x\prec u \land u\prec y)\lor(y\prec u\land u\prec z)\big)\\
&&\ \ \ \to ddist(x,y,z,A_2) \big)\Big)
\end{eqnarray*}\vspace{-3 pt}

\noindent In words, we select a set $Q$ from $A_2$ so that for any three consecutive
selected elements the distance from the second to the third is twice as large
as the distance from the first to the second. The elements $f,s,l$ are the
first, second and last element of $Q$ respectively. We make sure to select the
first two consecutive elements of $Q$ ($f$ and $s$) so that the first distance
is 1.  The total size of $A_2$ must then be
$(1+2+4+\ldots+2^{|A_1|-1})+1=2^{|A_1|}$, where the sum is obtained by adding
the consecutive distances, and we add one at the end because the last element
$l$ was not counted.

Finally, we will need the following MSO formulas: $root_L(A_1,A_2)$ which
checks if $|A_2|=|A_1|^2$, assuming  $|A_1|\le L$; $div_L(A_1,A_2)$ which
checks if $|A_1|$ divides $|A_2|$ assuming $|A_1|\le L$; and $mod_L(A_1,A_2,R)$
which is true if $|A_2|\bmod |A_1| = |R|$ and $|A_1|\le L$.  Let us also give
details for these.\vspace{-3 pt}

\begin{eqnarray*}
root_L(A_1,A_2) &:=& \exists Q \Big( Q\subseteq A_2 \land \forall S \big( section(S,A_2,Q) \to eq_L(S,A_1)\big) \land\\
&& \exists S' \big( S'\subseteq A_2 \land eq_L(S',A_1) \land \forall S (section(S,P_2,Q) \to (|S\cap S'|=1))\big)\Big)
\end{eqnarray*}\vspace{-3 pt}

\noindent Informally, we can divide $A_2$ into sections of size
$|A_1|$ and if we select a set $S'$ that contains exactly one
representative from each section then $|S'|=|A_1|$.\vspace{-3 pt}

\begin{eqnarray*}
div_L(A_1,A_2) &:=& \exists Q \big( Q \subseteq A_2\land \forall S\big( section(S,A_2,Q) \to eq_L(S,A_1))\big)\\
less_L(A_1,A_2) &:=& \exists S\big( (A_1\subset S)\land eq_L( S,A_2)\big) \\
mod_L(A_1,A_2,R) &:=& \exists S \big( S \subseteq A_2 \land eq_L(S,R) \land div_L(A_1,A_2\setminus S) \land less_L(R,A_1)\big)
\end{eqnarray*}\vspace{-6 pt}

\noindent The $div_L$ formula decides if $|A_1|$ exactly divides
$|A_2|$ by partitioning $A_2$ into sections of size $|A_1|$. Using
this we can then calculate remainders.

Finally, let us define a predicate that allows us to talk about binary numbers.
The formula $bit_L(A_1,A_2)$ will be true if $|A_1|$ is at most $L$ and, when
$|A_1|$ is written in binary, the bit in position $|A_2|$ is 1. Here, we number
the bits in the binary representation of $|A_1|$ so that the least significant
bit is in position 0.\vspace{-3 pt}

\begin{eqnarray*}
bit_L(A_1,A_2) &:=& \exists A_3\exists A_4 \Big( exp_L(A_2,A_3) \land double(A_3,A_4) \land \\
&& \exists R \big( mod_L(A_1,A_4,R) \land \neg less_L(R,A_3) \big) \Big)
\end{eqnarray*}\vspace{-6 pt}

\noindent Let us explain this informally. To check if the bit at
position $i=|A_2|$ is 1, we construct a set $A_3$ of size $2^i$ and a
set $A_4$ of size $2^{i+1}$. Then, we use $mod_L$ to construct a set
$R$ whose size is equal to $|A_1|\bmod 2^{i+1}$. If this remainder set
is at least as large as $A_3$ the bit in position $i$ is 1.

\subsection{Hardness for Unary Strings and Paths}

\begin{thm} \label{thm:paths}

Let $f$ be an elementary function and $c$ a constant. If there exists an
algorithm which, given a unary string $w$ of length $n$ and an MSO formula
$\phi$ decides if $w\models \phi$ in time $f(|\phi|)n^c$ then E=NE.

\end{thm}

\proof

Suppose that we are given a non-deterministic Turing machine that runs in time
$2^{kn}$, for some constant $k$, when given $n$ bits of input. We will use the
hypothetical algorithm to decide whether the machine accepts an arbitrary input
in deterministic exponential time. 

Let us discuss some technical details about the machine. Without loss of
generality, assume that we are given a non-deterministic machine that always
terminates in time $T= 2^{kn}$.  Assume that the machine uses a binary
alphabet, and without loss of generality it never uses more than $T$ cells of
tape. Also without loss of generality, we may assume that the first thing the
machine does is to non-deterministically guess a string of bits and use it to
fill out its tape. From that point on the machine behaves deterministically,
that is, there is a finite set of states $Q$ and a transition function $\delta:
Q\times\{0,1\} \to Q\times\{0,1\}\times\{L,S,R\}$, that tells the machine for
each state and cell contents, which state to go to next, what to write on the
current cell and whether to move left, right, or stay at the same cell.  The
state set $Q$ contains a special state $q_{acc}$ such that if the machine ever
enters this state it automatically accepts and never leaves this state.

Suppose that we have been given the description of such a machine with $|Q|$
states, where $|Q|$ is independent of the input, and $n$ bits of input. We will
construct a unary string $w$ of appropriate length and an MSO formula $\phi$
such that $w\models \phi$ if and only if the machine would accept this input.

Let $I$ be the number whose binary representation is exactly the input given to
the machine (so $I\le 2^n$). Construct a unary string $w$ of length
$L=(2I+1)T^2$, where we recall that $T$ is the upper bound on the machine's
running time. Now we need to construct the formula $\phi$.

Rather than giving all formal details, we will now give a high-level
description of $\phi$ and the reader may verify that $\phi$ can indeed be
constructed using the tools from the previous section. Our formula will first
ensure the following:

\begin{itemize} 

\item It will identify a subset of the input of size $I$ and another with size
$T^2$. This is achievable by observing that the largest odd divisor of $L$ is
$2I+1$, so we simply ask for the largest odd set whose size exactly divides the
input.

\item Using the root formula we partition the set of size $T^2$ into $T$
equally-sized sections. Each will correspond to a snapshot of the tape during a
step in the machine's execution.

\item Identify the first section of the tape, which will have length
$T=2^{kn}$. Then identify a prefix of it of size $kn$ (this can be done with
the $exp$ formula). Identify a prefix of that with size $n$. To do this, we
first identify a set of size $k$ (since $k$ is a fixed constant, this can be
done with a constant-size formula). We then use $div$ to identify a set of size
$n$ and then use $eq$ to find a prefix of size $n$.  This is where the
machine's input will initially be stored. 

\end{itemize}

\noindent It should be clear that the above can be expressed in MSO
with the formulas of the previous section. So at this point, we have
identified $T$ sections, each of size $T$, to represent the machine's
tape. Specifically, each section is a snapshot of the machine's tape
at a particular point in time. Each element thus naturally corresponds
to a specific cell at a specific point in time during the machine's
execution.

We also identify a special part at the start of the first tape section with
size $n$, where we will check that the input is stored, and a set of length $I$
whose size encodes the input.  Now we ask for the existence of a subset $B$ of
elements that will denote the cells of the tape where 1 is written. We also ask
for the existence of $|Q|$ sets, call them $H_i,\ i\in Q$.  The intended
meaning is that if a certain element from one of the tape sections is in $H_i$,
then the machine was in state $i$ at the point in time corresponding to that
section and the machine's head was located at the cell corresponding to that
element.

Once the above sets have been selected all of the machine's computation has
been encoded. Then we will just need to check that it's correct and accepting.
We thus express the following conditions:

\begin{itemize}

\item Ensure the input is correctly encoded at the start of the tape. To check
the bit at position $i$ we observe that the contiguous subset of the tape from
the beginning to that bit has size $i$. We can now use the $bit_L$ predicate on
the set $I$ to check that the correct digit is stored in each input position.

\item Ensure that the machine transitions correctly. We look at pairs of
elements that correspond to the same tape cell in consecutive steps in time,
that is, tape elements whose distance is exactly $T$, which can be verified
with the $eq_L$ formula. If the first has no $H_i$ label then either both have
$B$ or neither does. If the first has an $H_i$ label we check that the $B$
label changes appropriately for the other and an $H_j$ label is used
appropriately for the other or one of its neighbors, depending on the
transition function.

\item Finally, check that in each section of the tape exactly one element has
an $H_i$ label, and that it has exactly one. Also, check that some element
eventually gets the $H_{q_{acc}}$ label. 

\end{itemize}

\noindent All the above requirements can be checked with an MSO
formula with constant size (assuming $k,|Q|$ constant), except for the
use of the $eq_L$ predicate, which has size $2^{O(\log^*L)}$.  So the
whole formula also has size $2^{O(\log^*L)}$ and $L=2^{O(n)}$ so
$|\phi|=2^{O(\log^*n)}$.

Suppose that an algorithm with running time $f(|\phi|)|w|^c$ existed for
elementary $f$. Then, there exists $d$ such that $f(x)\le \exp^{(d)}(x)$.
Recall that $|w|=2^{O(n)}$.  So the running time is at most
$\exp^{(d+1)}(O(\log^*n)) 2^{O(n)}= 2^{O(n)}$.  \qed

\begin{cor} \label{cor:paths}

Let $f$ be an elementary function and $c$ a constant. If there exists an
algorithm which, given a path $P$ on $n$ vertices and an MSO formula $\phi$
decides if $P\models \phi$ in time $f(|\phi|)n^c$ then E=NE.

\end{cor}

\begin{cor} \label{cor:subdivisions}

Let $f$ be an elementary function, $c$ a constant, and $\mathcal{C}$ a
non-empty class of graphs closed under edge sub-divisions. If there exists an
algorithm which, given a graph $G\in\mathcal{C}$ on $n$ vertices and an MSO
formula $\phi$ decides if $G\models \phi$ in time $f(|\phi|)n^c$ then E=NE.
The same is true if $\mathcal{C}$ is closed under induced subgraphs and, for
all $d>0$, $\mathcal{C}$ contains a graph with diameter $d$.

\end{cor}

\proof

If the class is closed under induced subgraphs and for each $d>0$ there is a
graph in the class with diameter $d$, then the class contains all paths and
therefore we can invoke Corollary \ref{cor:paths}.  To see this, for each $d$
take the graph with diameter $d$ and let $u,v$ be two vertices with shortest
path distance $d$. The graph induced by $u,v$ and the vertices that make up a
shortest path from $u$ to $v$ is a path, since if more edges were induced a
shorter path would exist from $u$ to $v$.  Thus, the class contains a path with
$d+1$ vertices.

If the class is closed under edge subdivisions we can reduce (in fact
interpret) the MSO model checking problem on paths to MSO model checking on the
class. We have a path with $n$ vertices and by the proof of Theorem
\ref{thm:paths} we can assume $n$ to be even. Select the smallest graph in
$\mathcal{C}$, call it $G_m$, and subdivide its edges an appropriate number of
times so that all maximal connected sets of degree two vertices have odd size.
Select one such set and subdivide its edges so that it has size $n$ (this is
always possible if we started with a sufficiently large $n$). It is not hard to
amend the original formula so that it first locates this path of size $n$ that
we created (it's now the only maximal connected set of degree two vertices with
even size) and only works with vertices from it. Since we only subdivided edges
the graph we obtain is still in $\mathcal{C}$.  \qed

\subsection*{Further Consequences}

Let us now give two further applications of the ideas of Theorem
\ref{thm:paths}. First, we can extend these ideas to obtain an alternative,
self-contained proof of a result given in \cite{CourcelleMR00}: MSO$_2$
model-checking on cliques is not in XP, unless E=NE. In \cite{CourcelleMR00}
this is proved under the equivalent assumption P$_1\neq\textrm{NP}_1$ (the
P$\neq$NP assumption for unary languages).  That proof relies on the work of
Fagin on graph spectra \cite{Fagin74}.

Here we can simply reuse the ideas of Theorem \ref{thm:paths} by observing two
basic facts: first, with an appropriate MSO$_2$ formula we can select a set of
edges in the given clique that induces a spanning path. Therefore, we can
assume we have the same structure as in the case of paths. Second, the $eq_L$
predicate can be constructed in constant size, since two disjoint sets of
vertices are equal if and only if there exists a perfect matching between them
in the clique (and this is MSO$_2$-expressible).

\begin{cor} \label{cor:cliques}

If there exists an algorithm which, given a clique $K_n$ on $n$ vertices and an
MSO$_2$ formula $\phi$ decides if $K_n\models \phi$ in $n^{f(|\phi|)}$, for any
function $f$, then E=NE.

\end{cor}

\proof

The proof follows similar lines as in Theorem \ref{thm:paths}, so we only
explain here the differences. First, we must implement the $\prec$ predicate on
the clique. This is achieved by selecting a set of edges that induces a path
and then using the same tricks we used to simulate MSO for strings with MSO$_1$
for paths. Second, we must implement an $eq(A_1,A_2)$ predicate with constant
size. If we do both of these, the rest of the proof of Theorem \ref{thm:paths}
goes through unchanged, since the formula we construct only uses the $\prec$
and $=$ predicates and has constant size except for the $eq_L$ predicate.

The main observation is that for two sets of vertices $A_1,A_2$ it is possible
to express in MSO$_2$ logic the property ``there exists a perfect matching from
$A_1$ to $A_2$'', that is, there exists a set of edges such that all vertices
of $A_1\cup A_2$ are incident on a unique edge and all edges have exactly one
endpoint in $A_1$.  Let $pm(A_1,A_2)$ be a formula encoding the perfect
matching property. We can now define $eq$ as follows:\vspace{-3 pt}

\begin{eqnarray*}
eq(A_1,A_2) &:=& pm(A_1\setminus A_2,A_2\setminus A_1)
\end{eqnarray*}\vspace{-6 pt}

\noindent Thus, all that is left is to implement the $\prec$
predicate. We give here a high-level argument. First, we will ask for
the existence of a set $F$ of edges with the following properties:

\begin{itemize}

\item All vertices have exactly two edges of $F$ incident to them.

\item For any partition of the vertices there exists an edge from $F$ with
endpoints on both sides (connectivity).

\end{itemize}

\noindent It is not hard to see that $F$ induces a spanning cycle. Let
$F'$ be the set obtained from $F$ by removing an arbitrary edge and
let $s$ be one of the endpoints of the removed edge. We will now say
that $x\prec y$ if any subset of edges from $F'$ that connects $s$ to
$y$ must touch $x$.

The size of the clique we construct is the same as the length of the string in
Theorem \ref{thm:paths}. So if there exists an algorithm running in time
polynomial in the order of the input clique for fixed-size formulas then we can
simulate a NE Turing machine in deterministic exponential time and E=NE. \qed

Let us now take a look at the order-invariant and successor-invariant versions
of the problem. In this setting the problem is usually posed as follows (see
e.g. \cite{EngelmannKS12,EickmeyerKK13}): we are given a graph $G$ and an MSO
formula $\phi$ on graphs that also uses an ordering predicate $<$. We are
promised that $\phi$ is order-invariant. The question is if there exists an
ordering of the vertices $<$ such that $(G,<)\models \phi$. The observation now
is that, since the proof of Theorem \ref{thm:paths} only relies on the
existence of some arbitrary ordering (and uses a path to construct the
ordering) in the order-invariant case we can get the same hardness even for
edgeless graphs.

\begin{cor} \label{cor:order}

Let $f$ be an elementary function and $c$ a constant. If there exists an
algorithm which, given an edge-less graph $G$ on $n$ vertices and an
order-invariant MSO$_1$ formula $\phi$ decides if there exists an ordering $<$
of the vertices such that $(G,<)\models \phi$ in time $f(|\phi|)n^c$ then E=NE.

\end{cor}

\proof

Recall that we have established a similar hardness result in Theorem
\ref{thm:paths} for unary strings, in which case the formula constructed uses
only the predicate $\prec$. From a unary path of length $n$ we construct an
edge-less graph on $n$ vertices, while in the formula $\phi$ we replace the
$\prec$ predicate with $<$. It is not hard to see that the formula we construct
is indeed order-invariant (since it does not use the $E()$ predicate at all).
\qed

Let us also note that Corollary \ref{cor:order} also applies to
successor-invariant formulas, as they are equivalent to order-invariant
formulas for MSO logic. 

\section{Tree-Depth}

In this section we give a lower bound result that applies to the model-checking
algorithm for trees of bounded height given by Gajarsk{\'y} and
Hlin{\v{e}}n{\'y} \cite{GajarskyH12}. Recall that a rooted $t$-colored tree is
a structure for which we are supplied a $C(x,y)$ predicate (encoding that $x$
is a child of $y$) and a set of unary predicates $P_i$. We recall here the main
result of \cite{GajarskyH12}:

\begin{thm}[\cite{GajarskyH12}] \label{thm:Gajarsky}

Let $T$ be a rooted $t$-colored tree of height $h \ge 1$, and let $\phi$ be an
MSO sentence with $r$ quantifiers. Then $T \models \phi$ can be decided by an
FPT algorithm in time $O\left( \exp^{(h+1)} \left(2^{h+5}  r(t + r)\right) + |V
(T)|\right)$.  

\end{thm}

Theorem \ref{thm:Gajarsky} is the main algorithmic tool used to obtain the
recent elementary model-checking algorithms for tree-depth and shrub-depth
given in \cite{GajarskyH12} and \cite{GanianHNOMR12}, since in both cases the
strategy is to interpret the graph into a colored tree of bounded height.

The running time given in Theorem \ref{thm:Gajarsky} is an elementary function
of the formula $\phi$, but non-elementary in the height of the tree. Though we
would very much like to avoid that, it is not hard to see that the dependence
on at least one of the parameters must be non-elementary, since allowing $h$ to
grow eventually gives the class of all trees so the lower bound result of Frick
and Grohe should apply.

It is less obvious however what the height of the exponentiation tower has to
be exactly, as a function of $h$, the height of the tree. The fact that we know
that the height of the tower must be unbounded (so that we eventually get a
non-elementary function) does not preclude an algorithm that runs in time
$\exp^{(\sqrt{h})}(|\phi|)$ or, less ambitiously, $\exp^{(h/2)}(|\phi|)$, or
even $\exp^{(h-5)}(|\phi|)$. Recall that we are trying to determine the number
of levels of exponentiation in the running time here, so shaving off even an
additive constant would be a non-negligible improvement.

We show that even such an improvement is probably impossible, and Theorem
\ref{thm:Gajarsky} determines precisely the complexity of MSO model-checking on
colored trees of height $h$, at least in the sense that it gives exactly the
correct level of exponentiations for constant $h$. We establish this fact
assuming the ETH, by combining lower bound ideas which have appeared in
\cite{FrickG04} and \cite{Lampis12}. The main technical obstacle is comparing
indices, or in other words, counting economically in our construction. For
this, we use the tree representation of numbers of \cite{FrickG04} pruned to
height $h-1$. We then use roughly $\log^{(h)}n$ colors to differentiate the
leaves of the constructed trees.

The basic idea of our reduction is to start from an instance of $n$-variable
3SAT and construct an instance made up of a tree of height $h$ colored with
$t=O(\log^{(h)}n)$ colors. The formula will use $O(1)$ quantifiers, so the
algorithm of Theorem \ref{thm:Gajarsky} would run in roughly
$\exp^{(h+1)}(O(\log^{(h)}n))$ time. If an algorithm running in
$\exp^{(h+1)}(o(\log^{(h)}n))$ time existed we would be able to obtain a
$2^{o(n)}$ algorithm for 3SAT. Thus, the algorithm is optimal up to the
constant factor in the final exponent.

\begin{thm} \label{thm:treecount}

If for some constant $h\ge 1$ there exists a model-checking algorithm for
$t$-colored rooted trees of height $h$ that runs in $\exp^{(h+1)}(o(t))\cdot
poly(n)$ time for trees with $n$ vertices then the Exponential Time Hypothesis
fails.

\end{thm}

\begin{figure}

\centering

\includegraphics[width=0.8\textwidth]{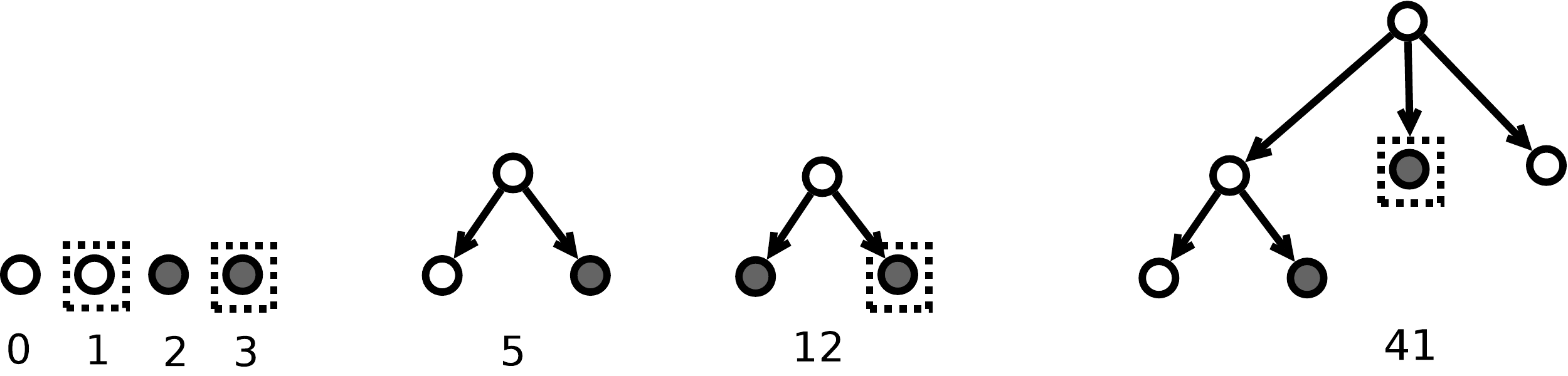}

\caption{An example of the graphs constructed in the proof of Theorem
\ref{thm:treecount}. Assume we have two labels available, one represented by
solid grey fillings and the other with a dashed box around vertices that have
it. The first 4 numbers ($0,\ldots,3$) can be represented by a single vertex.
Numbers up to $2^4-1$ can be represented with trees of height 1, numbers up to
$2^{2^4}-1$ with trees of height 2, etc. } \label{fig:treecounting}

\end{figure}

\proof

As usual in such proofs, the main obstacle is how to encode numbers up to $n$
economically in terms of the height of the constructed tree and the number of
colors used.  Fix some $h\ge 2$ (we will handle the case $h=1$ at the end). We
have at our disposal around $\log^{(h)}n$ colors. By using them we can create
$2^{\log^{(h)}n}=\log^{(h-1)}n$ vertices which we can distinguish by using a
different set of colors for each vertex. To go from there to $n$ we will use
the trick of \cite{FrickG04} which, roughly speaking, gives exponentially more
counting power with each level of height added. Thus, we will manage to
represent numbers up to $n$ with trees of height $h-1$.

Let us now be more precise. We have at our disposal $\log^{(h)}n$ colors,
number them $0,\ldots,\log^{(h)}n-1$. We will define for each
$i\in\{0,\ldots,n-1\}$ a rooted colored tree $T_i$. The construction is
inductive:

\begin{itemize}

\item If $i\in\{0,\ldots,\log^{(h-1)}n-1\}$ then $i$ has a binary
representation with at most $\log^{(h)}n$ bits, say $b_kb_{k-1}\ldots b_1b_0$
with $k\le \log^{(h)}n-1$. The tree $T_i$ is a single vertex colored with
exactly the colors $j$ such that $b_j=1$.

\item Suppose that we have defined $T_i$ for $i\in\{0,\ldots,\log^{(k)}n-1\}$
for some $k\ge 1$. We will now define $T_i$ for $\log^{(k)}n \le i \le
\log^{(k-1)}n-1$. As previously, write down the binary representation of $i$,
which has at most $\log^{(k)}n$ bits. For each $j$ such that $b_j=1$ construct
a copy of $T_j$ (we already know how to do this by the inductive hypothesis).
Add a new vertex, which will be the root of the new tree, and connect it to the
roots of the constructed trees. 

\end{itemize}

\noindent Some examples of the above construction are given in Figure
\ref{fig:treecounting}. Now let $i$ be an integer such that $1\le i\le
h$. We observe that the above construction represents numbers which
are at most $\le \log^{(h-i)}n-1$ with trees of height $i-1$. This can
be proved by induction: for $i=1$ trees of height $0$ (that is, single
vertices) are used to represent numbers up to $\log^{(h-1)}n-1$. For
the inductive case, notice that each level of height added increases
the maximum number representable exponentially. As a result, the
numbers $0,\ldots,n-1$ can be represented with a tree of height $h-1$.

We can now also define an $eq_k(x,y)$ predicate, that will be true if and only
if $x,y$ are the roots of two trees of height at most $k$ representing the same
number. Again we proceed inductively:

\begin{itemize}

\item It's easy to define a simple propositional predicate
$\mathrm{samecols}(x,y)$. The predicate will be true if $x,y$ have exactly the
same colors from the set $\{0,1,\ldots,\log^{(h)}n-1\}$.  Using this, for $k=0$
we set $eq_0(x,y):=\mathrm{samecols}(x,y) \land \forall z (\neg C(z,x) \land
\neg C(z,y))$. In other words, $x,y$ are equal if they have the same colors and
no children. 

\item Suppose $eq_k(x,y)$ is defined, we will define
  $eq_{k+1}(x,y)$. We set\vspace{-3 pt} 

\begin{eqnarray*} 
eq_{k+1}(x,y) &:=& \mathrm{samecols}(x,y) \land \forall u
\Big( \left( C(u,x) \lor C(u,y)\right) \to\\
&&\exists v \Big(eq_k(u,v)\land\\
&&\ \ (C(u,x)\to C(v,y)) \land (C(u,y)\to C(v,x))\Big) \Big)
\end{eqnarray*}\vspace{-6 pt}

\noindent In words, $x,y$ have the same colors and for every vertex
that is the child of one of them there exists a vertex that is a child
of the other and these two vertices represent the same number.

\end{itemize}

\noindent It is not hard to see that the formula $eq_h(x,y)$ uses
$O(h)$ quantifiers. We are now ready to describe our construction.

Fix $h>1$ and start with an instance of 3SAT with $n$ variables and suppose
that these variables are named $x_i$, $i\in\{n,\ldots,2n-1\}$. The reason we
number the variables this way is that it will be convenient for all of them to
have an index high enough that a non-trivial tree is needed to describe it. For
each variable $x_i$ construct a copy of the tree $T_i$ as described above. We
will make use of $\log^{(h)}2n = \log^{(h)}n+o(1)$ colors so trees have height
at most $h-1$. Color the roots of all these trees with a new color, call it
$v$.

For each clause $(l_i \lor l_j \lor l_k)$ where $l_i,l_j,l_k$ are literals,
that is, positive or negative appearances of the variables $x_i,x_j,x_k$
respectively, construct three trees $T_i, T_j, T_k$. Introduce six new colors,
call them $c_{p,q}$ for $p\in [3], q\in\{0,1\}$. If the literal $l_i$ is
positive then color the children of the root of the tree $T_i$ with $c_{1,1}$,
otherwise color them with $c_{1,0}$. Similarly, color the children of the root
of $T_j$ with $c_{2,1}$ if $l_j$ is positive and $c_{2,0}$ if it's negative and
the children of the root of $T_k$ with $c_{3,1}$ or $c_{3,0}$.  Notice that we
know that all trees have height at least 1 because $h\ge 2$ and variables are
numbered $n,\ldots,2n-1$, so all roots do have children. Finally, merge the
roots of $T_i, T_j, T_k$ into a single vertex. We introduce a new color, call
it $c$, and use it to color the new root of the tree that represents each
clause. To complete the construction, add a new root vertex to the graph and
make all roots of previously constructed trees its children. This creates a
tree with height $h$. The root has one $v$-colored child for each variable and
one $c$-colored child for each clause of the 3SAT formula.

Now, let us describe a formula with $O(h)$ quantifiers that will check if the
original 3SAT instance was satisfiable. Informally, we will ask if there exists
a set of variables, represented by a subset of the vertices colored with $v$,
such that setting these to true and the rest to false satisfies the formula. To
do this, we need to be able to check if a variable appears positive or negative
in a clause. Let's define a predicate $pos_i(x,y)$ which will be true if
variable $x$ appears positive in position $i$ (where $i\in [3]$) in the clause
represented by $y$ (so $x$ is assumed to be the root of a variable tree and $y$
is assumed to be the root of a clause tree).\vspace{-3 pt}

\begin{eqnarray*}
pos_i(x,y) &:=& \forall u \Big(\left(C(u,x) \lor (C(u,y)\land P_{c_{i,1}}(u))\right)\to \\
&& \exists v\ \Big(eq_{h}(u,v) \land\\
&&\ \ (C(u,x) \to (C(v,y)\land P_{c_{i,1}}(y))) \land (C(u,y)\to C(v,x))\Big)\Big)
\end{eqnarray*}\vspace{-6 pt}

\noindent The logic here is exactly the same as in the $eq$ predicate,
except that we only take into account the children of the clause node
that correspond to the $i$-th literal. It's easy to see how to make a
similar predicate $neg_i$ for negative appearances (change
$P_{c_{i,1}}$ to $P_{c_{i,0}}$). Given these, the complete formula
is:\vspace{-3 pt}

\begin{eqnarray*}
isSAT &:=& \exists S (\forall x (x\in S \to P_v(x))) \land \\
&& \forall y (P_c(y)\to \exists z (P_v(z)\land\\
&& (\lor_{i\in [3]} (pos_i(z,y)\land z\in S)) \lor   (\lor_{i\in [3]}
(neg_i(z,y)\land \neg z\in S))  ))
\end{eqnarray*}\vspace{-6 pt}

\noindent In words, there exists a set of variables $S$ (which will be
set to true), such that for each clause there exists a variable
appearing in it that satisfies it, that is, it belongs in $S$ if and
only if its appearance is positive.

The formula has $h$ quantifiers, we have used $\log^{(h)}n+O(1)$ colors and
have constructed a tree of height $h$ and size polynomial in $n$. If there
exists a model-checking algorithm for $t$-colored trees running in
$\exp^{(h+1)}(o(t)) |V|^c$ this gives a $2^{o(n)}$ algorithm for 3-SAT.

The only thing left is the case $h=1$. Here each variable and each clause will
be represented by a single vertex and, since we have $O(\log n)$ colors
available, the colors alone will be sufficient to compare indices. It's not
hard to see how to encode the whole structure of the formula using $7\log n$
colors. The first $\log n$ colors are used for the variable vertices. Then we
need $6$ sets of $\log n$ distinct colors to encode the appearances of literals
into the clauses, for each combination of position and positivity. This makes
it straightforward to implement $pos_i$ and $neg_i$ by comparing appropriate
sets of colors on the two vertices. The rest of the formula is unchanged.\qed

\section{Conclusions and Open Problems}

We have proved non-elementary lower bounds for FO logic on cographs and MSO
logic on uncolored paths. The hope is that, since these lower bounds concern
very simple graph families, they can be used as ``sanity checks'' guiding the
design of future graph widths. We have also given a lower bound for MSO logic
on colored trees of bounded height. It would be interesting to see if this can
be extended to uncolored trees.

Finally, let us mention that a promising direction in this area that we did not
tackle here is that of alternative logics, besides FO and MSO variants. One
example is the meta-theorems given by Pilipczuk \cite{Pilipczuk11} for a kind
of modal logic. The algorithmic properties of such logics are still mostly
unexplored but they may be a good way to evade the lower bounds given in
\cite{FrickG04} and this paper.

\noindent \textbf{Acknowledgement:} I am grateful to an anonymous reviewer for
pointing out that Theorem \ref{thm:paths} can be established using the weaker
complexity assumption E$\neq$ NE. A previous version of this paper used the
assumption that EXP$\neq$ NEXP.

\bibliographystyle{abbrv} 

\bibliography{biblio}

\end{document}